\documentstyle[aps,prd,amsfonts,amssymb,eqsecnum,preprint]{revtex}
\tighten

\newcommand{\real}{{\text{I\kern-.2em R}}}

\begin{document}

\title{%
  Conserved masses in {\sc ghs} Einstein and string black holes\\
  and consistent thermodynamics}
\author{%
  K. C. K. Chan,%
  \thanks{Electronic address:
    {\tt kckchan{\char'100}avatar.uwaterloo.ca}}
  J. D. E. Creighton,%
  \thanks{Electronic address:
    {\tt jolien{\char'100}avatar.uwaterloo.ca}}
  and R. B. Mann%
  \thanks{Electronic address:
    {\tt mann{\char'100}avatar.uwaterloo.ca}}}
\address{Dept.~of Physics, University of Waterloo, Waterloo, Ontario,
  Canada.  N2L~3G1}
\date{29 April, 1996}
\maketitle

\begin{abstract}\noindent
We analyze the relationship between quasilocal masses calculated for
solutions of conformally related theories.
We show that the {\sc adm} mass of a static, spherically
symmetric solution is conformally invariant (up to a constant factor)
only if the background action functional is conformally invariant.
Thus, the requirement of conformal invariance
places restrictions on the choice of reference spacetimes.
We calculate the mass of the black hole solutions obtained by
Garfinkle, Horowitz, and Strominger ({\sc ghs}) for both the string
and the Einstein metrics.  In addition, the quasilocal thermodynamic
quantities in the string metrics are computed and discussed.
\end{abstract}

\narrowtext
\section{Introduction}
\label{sec:intro}

General considerations regarding the structure of quantum gravity
suggest that general relativity may require some modification,
even at energies considerably lower than the Planck scale.
Dilatonic gravity theories have recently played a prominent role
in these considerations since dilaton gravity emerges as the
low-energy effective
field theory limit of string theories.  The general form of the
action of such theories is
\begin{equation}
  I = \int_{\cal M}{}^n\!\bbox{\epsilon}\, \bigl( D(\phi)R
  + H(\phi)(\nabla\phi)^2 + V(\phi) \bigr)
  + I_{\scriptscriptstyle\text{M}}
  \label{basic action}
\end{equation}
where $\cal M$ is a $n$-dimensional manifold with volume $n$-form
${}^n\!\bbox{\epsilon}$; $D$, $H$, and~$V$ are arbitrary functions of
the dilaton field~$\phi$; $R$ is the Ricci curvature scalar;
and~$I_{\scriptscriptstyle\text{M}}$ is the matter action.
In general, the matter action will be a functional
of the dilaton, the metric, and whatever matter fields are relevant
to the system of interest.

Specification of $D$, $H$, and~$V$ is tantamount to the specification
of a given dilaton theory of gravity in the absence of coupling to
matter.  When~$D$ does not depend upon the dilaton, the theory
represents general relativity with a minimally coupled scalar field.
However, we may always perform a conformal tranformation,
\begin{equation}
  g_{ab} = {\mit\Omega}^2(\phi) g^{\scriptscriptstyle\text{E}}_{ab}\;,
  \label{conformal transformation Einstein metric}
\end{equation}
that will give a metric that is a solution to the field equations
obtained from the action of eq.~(\ref{basic action})
with~$D(\phi)\propto{\mit\Omega}^{2-n}(\phi)$.  In
eq.~(\ref{conformal transformation Einstein metric}), the metric
representing a solution to Einstein's general relativity is denoted
by~$g^{\scriptscriptstyle\text{E}}_{ab}$.

It is often assumed that physical quantities, such as the mass and
thermodynamic variables, should be the same for conformally related
solutions.  For example, in a previous paper~\cite{GHS91}, Garfinkle,
Horowitz, and Strominger ({\sc ghs}) derived the $(3+1)$-dimensional
magnetically (or electrically) charged string black hole solutions
(for which $H/4=D\propto\exp(-2\phi)$) in terms of the Einstein metric
by performing the conformal tranformation
\begin{equation}
  g^{\scriptscriptstyle\text{S}}_{ab} = \text{e}^{2\phi}
  g^{\scriptscriptstyle\text{E}}_{ab}\;.
  \label{string conformal transformation}
\end{equation}
By assuming that a conformal transformation would preserve physical
quantities~\cite{H92} such as the {\sc adm} mass, they claimed that
the {\sc adm} mass in the string metric is the same as the one in the
Einstein metric.%
\footnote{By {\sc adm} mass, we mean the value of the Hamiltonian
conjugate to unit time translations at spacelike infinity for
asymptotically flat spacetimes.}

In this paper, we show that the situation is more subtle.
Specifically, we show that the quasilocal generalization of the
{\sc adm} mass for systems of finite size~\cite{BY93,CM95a} is not
generally conformally invariant.  This result is a consequence of
the lack of conformal invariance of the background action functional
that specifies a reference spacetime.  The mass is conformally
invariant only if the background action functional, which specifies
the reference spacetime, is required to be conformally invariant.
Even then the {\sc adm} mass may differ between conformally related
theories if the conformal transformation does not approach unity at
spacelike infinity, but this difference is just a constant factor due
to a trivial rescaling of the time coordinate.  As an illustration,
we use the {\sc ghs} string black hole solutions:  We calculate
the {\sc adm} mass directly from the string action using the
aforementioned quasilocal formalism~\cite{CM95a} as well as from the
Einstein action obtained via the conformal transformation of
eq.~(\ref{string conformal transformation}).  We show that the
{\sc adm} masses are the same, up to a constant factor, when a
suitable background action functional is chosen.  In addition, we
shall compute the thermodynamic quantities relevant to the string
metric and show that these are consistent with the first law of
thermodynamics.  It should be noted that the validity of the first law
of thermodynamics does not depend on the choice of the reference
spacetime, although the individual thermodynamic variables do.

In Sec.~\ref{sec:conformal}, we discuss the general circumstance
under which conformal invariance of the mass can be expected to hold,
and we construct a conformally invariant expression for the quasilocal
mass for a static, spherically symmetric spacetime.  We review the
definition of the quasilocal thermodynamic variables in
Sec.~\ref{sec:thermo}.  The example of the {\sc ghs} solution is
explored in Sec.~\ref{sec:ghs} where we compute the {\sc adm} mass
in both the string and the Einstein forms of the theory.  We also
calculate the thermodynamic variables for both the electric and the
magnetic string black hole solutions, and explore their asymptotically
large-system-size behaviour.
In the case of the magnetic string solution, we need to
modify the quasilocal formalism to avoid the Dirac string singularity
in the Maxwell potential.  Finally, we summarize our results in the
concluding section.

\section{Conformal Transformations and Quasilocal Mass}
\label{sec:conformal}

We wish to obtain a notion of mass for solutions of theories of
dilaton gravity that has the property that is invariant under
conformal transformations of the solutions.  That is, given two
solutions of conformally related theories that are themselves
conformally related, we want to have a definition of mass that
yields the same value for both solutions.  In this section we
shall address some of the problems involved in the construction
of such a mass.  Let us first summarize what it means for two
theories to be conformally related.

\subsection{Conformally Related Theories}

Consider two theories of dilaton gravity with metrics $g_{ab}$
and~${\tilde g}_{ab}$, dilaton fields $\phi$ and~$\tilde\phi$,
and actions
\begin{mathletters}
\label{actions}
\begin{eqnarray}
  I &=& \int_{\cal M} {}^n\!\bbox{\epsilon}\, \bigl( D(\phi)R
  + H(\phi)(\nabla\phi)^2 + V(\phi) \bigr) \nonumber \\
  &&\quad -2\int_{\partial{\cal M}} {}^{(n-1)}\!\bbox{\epsilon}\,
  D(\phi){\mit\Theta} - I_{\scriptscriptstyle0}
  \label{action}
\end{eqnarray}
and
\begin{eqnarray}
  {\tilde I} &=& \int_{\cal M} {}^n\!\bbox{\tilde\epsilon}\,
  \bigl( {\tilde D}({\tilde\phi}){\tilde R}
  + {\tilde H}({\tilde\phi})(\nabla{\tilde\phi})^2
  + {\tilde V}({\tilde\phi}) \bigr) \nonumber\\
  &&\quad -2\int_{\partial{\cal M}}
  {}^{(n-1)}\!\bbox{\tilde\epsilon}\,
  {\tilde D}({\tilde\phi}){\tilde{\mit\Theta}}
  - {\tilde I}_{\scriptscriptstyle0} \;.
  \label{tilde action}
\end{eqnarray}
\end{mathletters}%
Here, $\partial{\cal M}$ is the boundary of~$\cal M$ with
normal~$n^a$, induced metric~$\gamma_{ab}=g_{ab}-n_an_b$, extrinsic
curvature~${\mit\Theta}_{ab}=-\frac{1}{2}\mathord{\text{\pounds}}_n
\gamma_{ab}$, and volume
element~${}^{(n-1)}\!\bbox{\epsilon}=n\cdot{}^n\!\bbox{\epsilon}$.
The background action~$I_{\scriptscriptstyle0}$ is a functional of
the fields on~$\partial{\cal M}$ alone; the background action does
not contribute to the field equations, but it specifies a reference
spacetime that effectively defines zero mass.
The quantities with a tilde are defined similarly.

We wish to determine the relationship between the functions of the
dilaton under the requirement that the two theories are related by
the conformal tranformation
\begin{equation}
  {\tilde g}_{ab} = {\mit\Omega}^2(\phi) g_{ab} \quad \text{and} \quad
  {\tilde\phi} = {\mit\Upsilon}(\phi) \;.
  \label{conformal transformation}
\end{equation}
Under this transformation we have
\begin{eqnarray}
  {}^n\!\bbox{\tilde\epsilon}\,{\tilde D}({\tilde\phi}){\tilde R}
  &=& {}^n\!\bbox{\epsilon}\,{\mit\Omega}^{n-2}
  {\tilde D}({\tilde\phi}) \bigl( R - 2(n-1)\nabla^2\log{\mit\Omega}
  \nonumber\\
  &&\quad -(n-2)(n-1)(\nabla{\mit\Omega})^2 \bigr)
  \label{tilde R}
\end{eqnarray}
and
\begin{eqnarray}
  {}^{(n-1)}\!\bbox{\tilde\epsilon}\,{\tilde D}({\tilde\phi})
  {\tilde{\mit\Theta}} &=& {}^{(n-1)}\!\bbox{\epsilon}\,
  {\mit\Omega}^{n-1}{\tilde D}({\tilde\phi}) \bigl( {\mit\Omega}^{-1}
  {\mit\Theta} \nonumber \\
  &&\quad - (n-1)n^a\nabla_a\log{\mit\Omega} \bigr) \;.
  \label{tilde Theta}
\end{eqnarray}
It is clear that a necessary condition for the actions of
Eqs.~(\ref{actions}) to transform each other
is~${\tilde D}({\tilde\phi})={\mit\Omega}^{2-n}(\phi)D(\phi)$.
With this relation, we find
\begin{eqnarray}
  && \int_{\tilde{\cal M}}{}^n\!\bbox{\tilde\epsilon}\,
  {\tilde D}({\tilde\phi}){\tilde R}
  - 2\int_{\partial{\tilde{\cal M}}}{}^{(n-1)}\!\bbox{\tilde\epsilon}
  \,{\tilde D}({\tilde\phi}){\tilde{\mit\Theta}}\hfil \nonumber\\
  &&\hfil= \int_{\cal M}{}^n\!\bbox{\epsilon} \bigl( D(\phi)R + \cdots
  \bigl) - 2\int_{\partial{\cal M}}{}^{(n-1)}\!\bbox{\epsilon}\,
  D(\phi){\mit\Theta}
  \label{transform action}
\end{eqnarray}
where we have integrated by parts on the second last term of the right
hand side of eq.~(\ref{tilde R}).  The boundary terms so produced are
exactly cancelled by the $\mit\Omega$-dependent terms from the
extrinsic curvatures in eq.~(\ref{tilde Theta}).

The terms in the ellipsis of eq.~(\ref{transform action}) are
quadratic in~$\nabla\log{\mit\Omega}$ and, therefore, are quadratic
in~$\nabla\phi$.  Upon examination of these terms, it is
straightforward to show that equivalence between the two theories
further requires
\begin{equation}
  \Biggl( \frac{\tilde H}{\tilde D} - \frac{n-1}{n-2} \biggl(
  \frac{{\tilde D}'}{\tilde D} \biggr)^2 \Biggr) ({\mit\Upsilon}')^2
  = \Biggl( \frac{H}{D} - \frac{n-1}{n-2} \biggl( \frac{D'}{D}
  \biggr)^2 \Biggr)
  \label{requirement 1}
\end{equation}
and
\begin{equation}
  {\tilde D}^{n/(2-n)} {\tilde V} = D^{n/(2-n)} V
  \label{requirement 2}
\end{equation}
where the prime denotes a derivative with respect to the functional
argument.  (For the $n=2$ version of these relations, see
Ref.~\cite{M93}.)  Hence, the actions of Eqs.~(\ref{actions})
transform into one another under the
transformation~(\ref{conformal transformation}), provided the
reference actions appropriately transform into each other.

\subsection{Quasilocal Mass}

Because conserved quantities such as the mass are only well defined
with reference to the background action~$I_{\scriptscriptstyle0}$,
the conserved quantities defined in a theory given by the action of
eq.~(\ref{action}) will be the same as those defined in a theory
given by the action of eq.~(\ref{tilde action}) only if the background
action functionals also transform into one another under the
transformation~(\ref{conformal transformation}).  Then the conformal
transformation is just a field redefinition under which physical
quantities should be invariant.  However, if the background action
functionals do not tranform into each other, then the field equations
will still be conformally related, but the conserved quantities
(and thermodynamic variables) 
calculated from the solutions will not agree in general.  

Thus, it is important to select the appropriate background action 
functionals. Clearly the choice~$I_{\scriptscriptstyle0}=0$ has the 
appropriate tranformation property and is insensitive to any special 
symmetries a given solution may have. However, the mass (and other
conserved quantities)  will not generally 
have a finite limit as the quasilocal system expands to infinite size.

Consider as another choice the background action functionals
\begin{mathletters}
\label{background actions}
\begin{equation}
  I_{\scriptscriptstyle0} = -2\int_{\cal T} N
  \bbox{d}t\wedge{}^{(n-2)}\!\bbox{\epsilon} D(\phi) k_{\text{flat}}
  \label{background action}
\end{equation}
and
\begin{equation}
  {\tilde I}_{\scriptscriptstyle0} = -2\int_{\cal T}
  {\tilde N}\bbox{d}t\wedge{}^{(n-2)}\!\bbox{\tilde\epsilon}
  {\tilde D}({\tilde\phi}) {\tilde k}_{\text{flat}}.
  \label{tilde background action}
\end{equation}
\end{mathletters}%
Here, the timelike boundary~$\cal T$ is chosen to contain the
orbits of the timelike Killing vector~$(\partial/\partial t)^a$.
The lapse function~$N=(u^a\nabla_at)^{-1}$ measures the normalization
of the unit normal~$u^a$ to the spacelike surfaces~${\mit\Sigma}_t$
of constant~$t$.  The quantity~$k_{\text{flat}}$ is the trace of
the extrinsic curvature of the
$(n-2)$-surface~${\cal T}\cap{\mit\Sigma}_t$, together with its
geometry~$\sigma_{ab}$, in flat Euclidean space of dimension~$n-1$.
Similar definitions apply to the variables $\tilde N$
and~${\tilde k}_{\text{flat}}$.  Conserved quantities will remain
invariant under~(\ref{conformal transformation})
provided~(\ref{background action}) transforms
into~(\ref{tilde background action}) under this transformation.

Let us consider a specific example:  Suppose we are interested in a
static, spherically symmetric ({\sc sss}) solution to a theory of
dilaton gravity given by eq.~(\ref{action}) of the form
\begin{equation}
  ds^2 = -N^2(r)\,dt^2 + \frac{dr^2}{f^2(r)} + r^2\,d\omega^2
  \label{sss metric}
\end{equation}
where $d\omega^2$ is the line element of a unit $(n-2)$-sphere.
A natural choice for the boundary, $\cal T$,
of the manifold is the history of a $(n-2)$-sphere
of constant radial coordinate~$r=r_{\scriptscriptstyle\text{B}}$.
(We will neglect the initial and final spacelike boundary pieces.)
A solution to the conformally related theory~(\ref{tilde action}) is
\begin{equation}
  d{\tilde s}^2 = -{\mit\Omega}^2 N^2(r)\,dt^2
  + \frac{dr^2}{{\mit\Omega}^{-2}f^2(r)}
  + {\mit\Omega}^2 r^2\,d\omega^2 \;.
  \label{tilde sss metric}
\end{equation}
We find that
$k_{\text{flat}}=-(n-2)/r_{\scriptscriptstyle\text{B}}$
and~${\tilde k}_{\text{flat}}=-(n-2)/({\mit\Omega}
r_{\scriptscriptstyle\text{B}})$.
Thus, we see that the boundary action functionals given
in~(\ref{background actions}) suitably transform into each other.

The quasilocal mass~\cite{BY93,BCM94} is a charge associated with a
timelike Killing vector on the boundary manifold~$\cal T$.  Unlike the
quasilocal energy defined below, the quasilocal mass is conserved in
the sense that it does not depend upon any foliation of the boundary
manifold~$\cal T$; in the absence of matter flux through~$\cal T$,
the value of the quasilocal mass does not change with time.  For a
{\sc sss} spacetime solution of the form~(\ref{sss metric}) to the
theory given by the action~(\ref{action}) with background
action~(\ref{background action}), the quasilocal mass is~\cite{CM95a}
\begin{eqnarray}
  M(r) = 2N(r) \biggl( && \frac{(n-2)A_{n-2}(r)D(\phi)}{r} \nonumber\\
  &&\quad - f(r)\,\frac{d}{dr}\bigl(A_{n-2}(r)D(\phi)\bigr) \biggr)
  \label{mass}
\end{eqnarray}
where $A_m(r)$ is the surface ``area'' of an $m$-sphere of radius~$r$:
\begin{equation}
  A_m(r) = \frac{(4\pi)^{m/2}{\mit\Gamma}(m/2)}{{\mit\Gamma}(m)}\,r^m
  \;.\label{area}
\end{equation}
The mass of a {\sc sss} solution to the theory given by the
action~(\ref{tilde action}) with background
action~(\ref{tilde background action}) is
\begin{eqnarray}
  {\tilde M}(r) = 2{\tilde N}(r) \biggl( &&
  \frac{(n-2){\tilde A}_{n-2}(r){\tilde D}({\tilde\phi})}%
    {{\mit\Omega}r} \nonumber\\
  &&\quad -{\tilde f}(r)\,
  \frac{d}{dr}\bigl({\tilde A}_{n-2}(r){\tilde D}({\tilde\phi})\bigr)
  \biggr) \;.
  \label{tilde mass}
\end{eqnarray}
It is easily verified that~${\tilde M}(r)=M(r)$ if the solutions are
related by the conformal tranformation of
eq.~(\ref{conformal transformation})
since~${\tilde A}_{n-2}(r)={\mit\Omega}^{n-2}A_{n-2}(r)$.

Although the background action of eq.~(\ref{background action})
has many attractive features, 
there are several limitations to the notion of mass given by
eq.~(\ref{mass}).  Because we chose the background action functional
to be related to the extrinsic curvature of the boundary metric
embedded in flat space, the definition is only suited to solutions
that are asymptotically flat.  However, the notion of mass can be
easily generalized by replacing~$k_{\text{flat}}$ with the
trace of the extrinsic curvature of the
$(n-2)$-surface~${\cal T}\cap{\mit\Sigma}_t$ embedded in a spacelike
slice of a {\sc sss} spacetime with appropriate asymptotic behavior
that is considered to be a reference spacetime (if such an embedding
is possible).  For spacetimes that are not spherically
symmetric, it is not clear whether the background action functional
of eq.~(\ref{background action}) will have the appropriate conformal
tranformation property.

The {\sc adm} mass is defined
as the value of the gravitational Hamiltonian evaluated on a surface
at spacelike infinity in an asymptotically flat spacetime.  That this
definition of {\sc adm} mass agrees with the original
definition~\cite{ADM62} has been demonstrated in
Refs.~\cite{IW94,HH95} for General Relativity.
A similar analysis shows that the
Hamiltonian for the gravitational action of eq.~(\ref{action})
is the same as the integral over a spacelike hypersurface of the
time-time component of the metric equation of motion.
However, for the {\sc sss} metric of
eq.~(\ref{sss metric}), there is a disagreement between the
quasilocal mass and the above notion of an {\sc adm} mass:
If the conformal transformation does
not approach unity in the asymptotically flat regime, the asymptotic
value of the lapse function in one solution will not be the same as
the asymptotic value of the lapse in the conformally related
solution.  In the {\sc adm} prescription, we would
renormalize the time variable (since the solution is static) so that
the lapse approaches unity as the radius approaches infinity.
This will cause the two {\sc adm} masses to differ by a constant
factor equal to the asymptotic value of the conformal factor.

Rather than choosing the background action functional of
eq.~(\ref{background action}), one may choose the background action
along the lines proposed by Hawking and Horowitz~\cite{HH95}:
the reference action functional~$I_{\scriptscriptstyle0}$ has the
same functional form as the rest of the action~$I$, but is a
functional of the variables $(g_{\text{ref}})_{ab}$
and~$\phi_{\text{ref}}$, which are independent of $g_{ab}$ and~$\phi$
except on the boundary.  On the boundary we require
$(\gamma_{\text{ref}})_{ab}\equiv\gamma_{ab}$
and~$\phi_{\text{ref}}\equiv\phi$.
The reference spacetime is the solution to the field equations
produced from the action~$I_{\scriptscriptstyle0}$.  Under this
prescription, the mass of the {\sc sss} solution~(\ref{sss metric})
is
\begin{equation}
  M(r) = 2N(r)\bigl(f_{\text{ref}}(r)-f(r)\bigr)\,
  \frac{d}{dr}\bigl( A_{n-2}(r)D(\phi) \bigr)
  \label{HH mass}
\end{equation}
where the reference spacetime solution has the same form as
eq.~(\ref{sss metric}).  Notice that
$N_{\text{ref}}\equiv N$ and~$\phi_{\text{ref}}\equiv\phi$
since these are fields on the boundary manifold, while
$f_{\text{ref}}$ is independent of~$f$.  Typically, Minkowski
spacetime will be a {\sc sss} solution to the field equations
generated by the reference action functional; if this is chosen
as the reference spacetime, then we have~$f_{\text{ref}}=1$ in
eq.~(\ref{HH mass}).  Although the reference action
functional~$I_{\scriptscriptstyle0}$ is conformally invariant
(provided the requirements of the previous subsection are satisfied),
the mass of eq.~(\ref{HH mass}) will not be conformally invariant
because one would naturally choose the {\em same\/} reference
solution---rather than the conformal transform of the reference
solution---in the conformally transformed theory.

Nevertheless, there seem to be situations in which the mass
defined by eq.~(\ref{HH mass}) is more attractive than the
conformally invariant mass of eq.~(\ref{mass}).  For example,
in the asymptotically anti-de\thinspace Sitter $(2+1)$-dimensional
black hole solution of Mart{\'\i}nez and Zanelli~\cite{MZ96},
the mass obtained from eq.~(\ref{HH mass}) is finite (and agrees
with the result of Mart{\'\i}nez and Zanelli) as~$r\to\infty$, while
the mass obtained from eq.~(\ref{mass}) linearly diverges
as~$r\to\infty$.  Transforming to the ``Einstein frame'' ({i.e.}, the
form of the action with constant~$D$), we find that the mass has the
same divergent behaviour,
as expected since~(\ref{mass}) was chosen to be conformally invariant.
In the Einstein frame, the two prescriptions are equivalent, so the
mass obtained from~(\ref{HH mass}) likewise diverges.

We have shown that the quasilocal mass of a {\sc sss} solution of
dilaton gravity is conformally invariant provided that a suitable
background action functional can be chosen.  However, we should note
that the quasilocal energy, representing the thermodynamic internal
energy of the gravitating system, cannot be conformally invariant
for the following reason:  although the entropy, $S$, and the surface
gravity of a black hole are conformally invariant, the temperature
of a black hole, $T$, as measured on the finite-sized quasilocal
surface, is {\em not\/} conformally invariant because of the presence
of the blue-shift factor.  Thus, the thermodynamic internal energy,
which must satisfy the first law of thermodynamics,
$T\,dS=dE+\text{work terms}$, cannot be conformally invariant either.
In fact, conformal invariance of thermodynamic quantities should not
be expected because these quantites are observer-dependent (although
the laws of thermodynamics are not);
we discuss the thermodynamic variables in the next section.

\section{Thermodynamic Variables}
\label{sec:thermo}

Consider a dilaton theory of gravity given by the action of
eq.~(\ref{action}) with an additional Maxwell-dilaton interaction:
\begin{equation}
  I_{\scriptscriptstyle\text{M}} = \int_{\cal M}{}^n\!\bbox{\epsilon}
  \,{\textstyle\frac{1}{4}}W(\phi){\frak F}^{ab}{\frak F}_{ab}
  \label{Maxwell action}
\end{equation}
where $\bbox{\frak F}=\bbox{d{\frak A}}$ is the two-form strength
of the Maxwell field with one-form potential~$\bbox{\frak A}$.
In Ref.~\cite{CM95b}, it was shown that the first law of
thermodynamics for classical {\sc sss} black hole solutions to such
a theory is
\begin{eqnarray}
  \delta S &=& \int_{\bar{\cal T}} N\bbox{d}\tau \wedge \bigl(
  \delta\bbox{E} + {\cal S}\,\delta{}^{(n-2)}\!\bbox{\epsilon}
  \nonumber\\
  &&\quad + \bbox{\mu}\,\delta\phi + {\frak V}\,\delta\bbox{\frak Q}
  + \bbox{\frak K}^a\,\delta{\frak W}_a \bigr) \;.
  \label{integral first law}
\end{eqnarray}
Here, $N=(u^a\nabla_a\tau)^{-1}$~represents the lapse function
associated with the foliation of the spacetime into spatial
hypersurfaces of constant Euclidean time~$\tau$ and timelike
normal~$u^a$.  The foliation is degenerate at the event horizon
of the black hole; the Euclidean manifold must be periodic in
time with period~$\Delta\tau=2\pi/\kappa_{\scriptscriptstyle\text{H}}$
($\kappa_{\scriptscriptstyle\text{H}}$~is the surface gravity of
the event horizon) in order to avoid a conical singularity at the
event horizon.  The manifold~$\bar{\cal T}$ is the periodic
history of a $(n-2)$-sphere quasilocal surface~$\cal B$ with volume
element~${}^{(n-2)}\bbox{\epsilon}$ and
radius~$r_{\scriptscriptstyle\text{B}}$.  $\bbox{E}$,
$\bbox{\mu}$, $\bbox{\frak Q}$, and~$\bbox{\frak K}^a$ are
$(n-2)$-forms representing the
quasilocal surface energy, dilaton potential, Maxwell charge, and
Maxwell current densities respectively.  $\cal S$ is the
quasilocal surface tension, while ${\frak V}=-u^a{\frak A}_a$
and~${\frak W}_a=\sigma^b_a{\frak A}_b$
with~$\sigma_{ab}=\gamma_{ab}+u_au_b$.  The entropy of the black hole
is given by
\begin{equation}
  S = 4\pi \int_{\cal H} {}^{(n-2)}\!\bbox{\epsilon}\, D(\phi)
    = 4\pi A_{n-2}(r_{\scriptscriptstyle\text{H}}) D(\phi)
  |_{\phi=\phi(r_{\text{H}})}
  \label{entropy}
\end{equation}
where $\cal H$ is the event horizon $(n-2)$-sphere with
radius~$r_{\scriptscriptstyle\text{H}}$,
and~$A_{n-2}(r_{\scriptscriptstyle\text{H}})$ is the surface ``area''
of the event horizon given by eq.~(\ref{area}).  Given the
conformal transformation properties of the area and the dilaton
function~$D(\phi)$ stated in the previous section, we see that the
entropy is conformally invariant.

Because the solution is static, the $\tau$ integral yields an overall
factor of~$\beta=N\Delta\tau$, which is the inverse temperature on the
quasilocal surface.  The quasilocal energy, dilaton potential, and
Maxwell charge can be calculated as follows~\cite{CM95b}:
\begin{eqnarray}
  E &= \int_{\cal B}\bbox{E} =& -2A
  \bigl( n^a\nabla_a D(\phi) - D(\phi)k \bigr)
  - E_{\scriptscriptstyle0}
  \label{energy}
\end{eqnarray}
\begin{eqnarray}
  \mu &= \int_{\cal B}\bbox{\mu} =& 2A \bigl( H(\phi)n^a\nabla_a\phi
  \nonumber \\
  &&\quad - (dD/d\phi)(k-n^a a_a) \bigr)
  - \mu_{\scriptscriptstyle0}
  \label{dilaton potential}
\end{eqnarray}
\begin{eqnarray}
  {\frak Q} &= \int_{\cal B}\bbox{\frak Q} =& -AW(\phi) n^a{\frak E}_a
  - {\frak Q}_{\scriptscriptstyle0} \;.
  \label{Maxwell charge}
\end{eqnarray}
Here, $a^b=u^a\nabla_au^b$
is the acceleration of the normal vector~$u^a$ (i.e., the acceleration
of an observer on the quasilocal surface),
$k_{ab}=-\frac{1}{2}\mathord{\text{\pounds}}_n\sigma_{ab}$ is the
extrinsic curvature of the quasilocal surface as embedded in a
spacelike leaf of the foliation, and~${\frak E}_a={\frak F}_{ab}u^b$
is the electric field strength.
$A=A_{n-2}(r_{\scriptscriptstyle\text{B}})$~is the ``area'' of the
$(n-2)$-sphere~$\cal B$ given by eq.~(\ref{area}).
The Maxwell surface current density, $\bbox{\frak K}^a$,
should not contribute to the first law of thermodynamics
in the spherically symmetric case as there is no preferred direction
on the quasilocal surface~$\cal B$.  However, we will need it in order
to calculate the electromagnetic work term for magnetic monopoles;
the surface current density is given by
\begin{equation}
  \bbox{\frak K}^a=-{}^{(n-2)}\!\bbox{\epsilon}\,W(\phi){\frak F}_{cd}
  \sigma^{ac}n^d - \bbox{\frak K}_{\scriptscriptstyle0}^a \;.
  \label{Maxwell current}
\end{equation}
The surface tension is given
by~\cite{CM95b,BCM94}
\begin{equation}
  {\cal S} =  2n^a\nabla_a D(\phi) + 2D(\phi) \Biggl(
  n^a a_a - k \biggl(\frac{n-3}{n-2}\biggr) \Biggr)
  - {\cal S}_{\scriptscriptstyle0} \;.
  \label{surface tension}
\end{equation}
Because of spherical symmetry, we can re-write
eq.~(\ref{integral first law}) in the differential form
\begin{equation}
  T\,dS = dE + {\cal S}\,dA + \mu\,d\phi + {\frak V}\,d{\frak Q}
  \label{first law}
\end{equation}
where $T=\beta^{-1}$ is the Hawking temperature blue-shifted to its
observed value on the quasilocal surface.

Each quantity in Eqs.~(\ref{energy})--(\ref{surface tension}) includes
a term (with a subscripted ``0'') that arises from the background
action functional~$I_{\scriptscriptstyle0}$.  We shall assume that
this action functional does not depend on the Maxwell field so that
${\frak Q}_{\scriptscriptstyle0}$
and~$\bbox{\frak K}_{\scriptscriptstyle0}^a$ both vanish.  We further
require the background action to be a linear functional of the
lapse~$N$ \cite{BY93}.  Then
\begin{equation}
  E_{\scriptscriptstyle0} = -\int_{\cal B}
  (\delta I_{\scriptscriptstyle0}/\delta N)
  \label{reference energy}
\end{equation}
is the reference energy~\cite{BY93}; the reference surface tension
and dilaton potential can be calculated from the reference energy via
${\cal S}_{\scriptscriptstyle0}=-(\partial E_{\scriptscriptstyle0}/
\partial A)$
and~$\mu_{\scriptscriptstyle0}=-(\partial E_{\scriptscriptstyle0}/
\partial\phi)$ \cite{CM95a}.

\section{The GHS Solution}
\label{sec:ghs}

The {\sc ghs} solution~\cite{GHS91,H92} in the ``string frame'' is a
solution of equations of motion arising from the action
\begin{eqnarray}
  I &=& \frac{1}{16\pi} \int_{\cal M} {}^4\!\bbox{\epsilon}\,
  \text{e}^{-2\phi} \bigl( R + 4(\nabla\phi)^2
  - {\frak F}^{ab}{\frak F}_{ab} \bigr) \nonumber \\
  &&\quad -\frac{1}{8\pi} \int_{\cal T} {}^3\!\bbox{\epsilon}\,
  \text{e}^{-2\phi}{\mit\Theta} - I_{\scriptscriptstyle0}\;.
  \label{ghs action}
\end{eqnarray}
We are interested in two solutions of the equations of motion
generated by this action: the electrically charged and the
magnetically charged black hole solution.  The magnetic string
black hole solution is~\cite{GHS91}
\begin{mathletters}
\label{magnetic ghs}
\begin{equation}
  \text{e}^{-2\phi} = \text{e}^{-2\phi_0} \biggl( 1 - \frac{q^2}{ar}
  \biggr)
  \label{magnetic phi}
\end{equation}
\begin{equation}
  \bbox{\frak F} = q\,\bbox{d}\vartheta \wedge \sin\vartheta\,
  \bbox{d}\varphi
  \label{magnetic F}
\end{equation}
\begin{equation}
  f^2(r) = \biggl( 1 - \frac{2a}{r} \biggr) \biggl( 1 - \frac{q^2}{ar}
  \biggl)
  \label{magnetic f2}
\end{equation}
and
\begin{equation}
  N(r) = f(r)\,\text{e}^{2(\phi-\phi_0)}
  \label{magnetic N}
\end{equation}
\end{mathletters}%
where the line element is given by eq.~(\ref{sss metric});
$\phi_{\scriptscriptstyle0}$, $q$, and~$a$ are constants of
integration.%
\footnote{The constant of integration, $a$, is the combination of
constants $m\text{e}^{\phi_0}$ used in Refs.~\cite{GHS91,H92}.}
This black hole has an event horizon
at~$r_{\scriptscriptstyle\text{H}}=2a$.
The electric string black hole solution is~\cite{H92}
\begin{mathletters}
\label{electric ghs}
\begin{equation}
  \text{e}^{-2\phi} = \text{e}^{2\phi_0} + \frac{q^2}{ar}
  \label{electric phi}
\end{equation}
\begin{equation}
  \bbox{\frak F} = -\frac{q\text{e}^{2\phi}}{r^2}\,
  N(r)\,\bbox{d}t \wedge f^{-1}(r)\,\bbox{d}r
  \label{electric F}
\end{equation}
\begin{equation}
  f^2(r) = 1 + \text{e}^{-2\phi_0} \biggl( \frac{q^2-2a^2}{ar} \biggr)
  \label{electric f2}
\end{equation}
and
\begin{equation}
  N(r) = f(r)\,\text{e}^{2(\phi+\phi_0)} \;.
  \label{electric N}
\end{equation}
\end{mathletters}%
The event horizon for this black hole solution is
at~$r_{\scriptscriptstyle\text{H}}=\text{e}^{-2\phi_0}(2a-q^2/a)$.
For both of these solutions, the quasilocal mass can be computed
with eq.~(\ref{mass}) if we assume that the appropriate choice of
the background action functional is made.  We find
$\lim_{r\to\infty}M(r)=a$~for the
electric solution and~$a\text{e}^{-2\phi_0}$ for the magnetic
solution.

Using the conformal
transformation~(\ref{string conformal transformation}) with the
conformal factor given by eq.~(\ref{magnetic phi}) and
eq.~(\ref{electric phi}) for the magnetic and electric solutions
respectively, we recover the solution in the ``Einstein frame.''
Let~$\varrho=r\text{e}^{-\phi}$; the metric in the Einstein frame
then yields
\begin{equation}
  ds^2 = -N^2\text{e}^{-2\phi}\,dt^2
  + \frac{d\varrho^2}{f^2\text{e}^{2\phi}(d\varrho/dr)^2}
  + \varrho^2\,d\omega^2
  \label{Einstein ghs metric}
\end{equation}
for either the magnetic or electric cases.  This is a solution of
the theory given by the action in the Einstein frame:
\begin{eqnarray}
  I &=& \frac{1}{16\pi}\int_{\cal M} {}^4\!\bbox{\epsilon}\, \bigl(
  R - 2(\nabla\phi)^2 - \text{e}^{-2\phi}{\frak F}^{ab}{\frak F}_{ab}
  \bigr) \nonumber\\
  &&\quad -\frac{1}{8\pi}\int_{\cal T} {}^3\!\bbox{\epsilon}\,
  {\mit\Theta} - I_{\scriptscriptstyle0} \;.
  \label{Einstein ghs action}
\end{eqnarray}
If the background action function is chosen appropriately, then
eq.~(\ref{mass}) can be used to calculate the {\sc adm} mass of the
solution~(\ref{Einstein ghs metric}) for both the electric and the
magnetic cases.  We find $\lim_{\varrho\to\infty}M(\varrho) = a$~for
the electric solution and~$a\text{e}^{-2\phi_0}$ for the magnetic
solution.  As expected, these masses agree with those we obtained from
the metric in the string frame.%
\footnote{In fact, the mass in the Einstein frame and in the
string frame agree for a quasilocal boundary at any radius, not
just at spacelike infinity.}  However,
when~$\phi_{\scriptscriptstyle0}\ne0$, the conformal transformation
does not approach unity at spatial infinity, and thus the lapse
function~$N\text{e}^{-\phi}$ of eq.~(\ref{Einstein ghs metric}) also
fails to approach unity at spatial infinity.  Since the solution is
static, we can rescale the time coordinate:
$t\to t_{\text{new}}=\text{e}^{\pm\phi_0}$ where the
plus (minus) applies to the magnetic (electric) solution.  The new
lapse function is~$N\text{e}^{-(\phi\mp\phi_0)}$.  The mass rescales
by a constant factor, and the
result~$M_{\scriptscriptstyle\text{ADM}}=a\text{e}^{-\phi_0}$ is
obtained for both the magnetic and electric solutions, in agreement
with Refs.~\cite{GHS91,H92}.  We note that such a rescaling is always
possible for static solutions and, unless there is some required
asymptotic behaviour for the lapse function, the definition of mass 
contains an arbitrary constant factor.

Consider now the mass defined by eq.~(\ref{HH mass}) with the
reference solution of Minkowski spacetime with a constant dilaton.
In the Einstein frame, the mass of eq.~(\ref{HH mass}) agrees with
the mass of eq.~(\ref{mass}) because~$D$ is constant.  However, in
the string frame, eq.~(\ref{HH mass}) yields
$M=\bigl(a-q^2/(2a)\bigr)$ for the electric solution
and~$M=\text{e}^{-2\phi_0}\bigl(a+q^2/(2a)\bigr)$ for the magnetic
solution.  Clearly the mass is not conformally invariant when
eq.~(\ref{HH mass}) is used.

\subsection{Thermodynamics of the Electric String Solution}

The electric string solution is given by Eqs.~(\ref{electric ghs}).
We can evaluate the extensive thermodynamic variables using
Eqs.~(\ref{entropy}), (\ref{energy}), and~(\ref{Maxwell charge}).
We find
\begin{equation}
  S = 2\pi a r_{\scriptscriptstyle\text{B}} \biggl(
  \frac{2a^2-q^2}%
    {ar_{\scriptscriptstyle\text{B}}\text{e}^{-2\phi}-q^2} \biggr)
  \label{electric entropy}
\end{equation}
is the entropy of the black hole,
\begin{equation}
  E = -f(r_{\scriptscriptstyle\text{B}})\,\biggl(
  r_{\scriptscriptstyle\text{B}}\text{e}^{-2\phi} - \frac{q^2}{2a}
  \biggr) - E_{\scriptscriptstyle0}
  \label{electric energy}
\end{equation}
is the quasilocal energy, and
\begin{equation}
  {\frak Q} = q
  \label{electric Maxwell charge}
\end{equation}
is the Maxwell charge.  Note that we can solve
eq.~(\ref{electric entropy}) for the parameter~$a$ to
obtain an expression for~$a$ in terms of the extensive variables
$S$, $\frak Q$, $\phi$, and~$A=4\pi r_{\scriptscriptstyle\text{B}}^2$.
We can write~$f(r_{\scriptscriptstyle\text{B}})$ in terms of these
extensive variables and, thus, write~$E=E(S,A,\phi,{\frak Q})$.  The
zero of the energy, $E_{\scriptscriptstyle0}$, is set by the
background action functional via~(\ref{reference energy}).

The intensive variables include the temperature, the surface tension,
the dilaton potential, and the Maxwell potential.  The temperature
is~$T=\kappa_{\scriptscriptstyle\text{H}}/(2\pi N)$ with
$\kappa_{\scriptscriptstyle\text{H}}=f(dN/dr)$ evaluated on the
event horizon.  We find
\begin{equation}
  T = \frac{1}{8\pi a N(r_{\scriptscriptstyle\text{B}})}\, \biggl(
  \text{e}^{-2\phi} - \frac{q^2}{ar_{\scriptscriptstyle\text{B}}}
  \biggr) \;.
  \label{electric temperature}
\end{equation}
Using eq.~(\ref{surface tension}), we compute the surface tension,
\begin{equation}
  {\cal S} = \frac{1}{8\pi r_{\scriptscriptstyle\text{B}}^2}\, \Biggl(
   f(r_{\scriptscriptstyle\text{B}})\, \biggl(
  r_{\scriptscriptstyle\text{B}}\text{e}^{-2\phi} - \frac{q^2}{2a}
  \biggr) + \frac{a}{f(r_{\scriptscriptstyle\text{B}})} \Biggr)
  - {\cal S}_{\scriptscriptstyle0}\;.
  \label{electric surface tension}
\end{equation}
We use eq.~(\ref{dilaton potential}) to calculate the dilaton
potential:
\begin{equation}
  \mu = -f(r_{\scriptscriptstyle\text{B}})\, \biggl(
  2r_{\scriptscriptstyle\text{B}}\text{e}^{-2\phi} - \frac{q^2}{2a}
  \biggr) - \frac{a}{f(r_{\scriptscriptstyle\text{B}})}
  - \mu_{\scriptscriptstyle0} \;.
  \label{electric dilaton potential}
\end{equation}
The contributions of the reference spacetime to the surface tension
and dilaton potential, ${\cal S}_{\scriptscriptstyle0}$
and~$\mu_{\scriptscriptstyle0}$, are determined in terms
of~$E_{\scriptscriptstyle0}$ as stated at the end of
Sec.~\ref{sec:thermo}.  Finally, the Maxwell potential, $\frak V$,
is given by~\cite{CM95b}
\begin{equation}
  {\frak V} = \frac{1}{N(r_{\scriptscriptstyle\text{B}})}
  \int^{r_{\text{B}}}_{r_{\text{H}}} {\frak F}_{tr}\, dr
  = - \frac{q}{2a}\,f(r_{\scriptscriptstyle\text{B}}) \;.
  \label{electric Maxwell potential}
\end{equation}
Notice that~$\frak V$ is finite on the event horizon.

We can express  the quasilocal energy (\ref{electric energy})
in terms of the extensive variables $S$, $A$, $\phi$, and $\frak Q$
using (\ref{electric entropy}) and (\ref{electric Maxwell charge}).
It can then be shown that
\begin{equation}
  T = \frac{\partial E}{\partial S}, \quad
  {\cal S} = -\frac{\partial E}{\partial A}, \quad
  \mu = -\frac{\partial E}{\partial\phi}, \quad\text{and}\quad
  {\frak V} = -\frac{\partial E}{\partial{\frak Q}}\;.
  \label{intensive variables}
\end{equation}
Thus, the quasilocal energy is the appropriate choice for the
thermodynamic internal energy of the gravitating system contained
within the quasilocal boundary; we have explicitly demonstrated the
first law of thermodynamics given by eq.~(\ref{first law}).  This
thermodynamic relation is independent of the choice of the background
action functional~$I_{\scriptscriptstyle0}$ and, thus, independent
of the choice of~$E_{\scriptscriptstyle0}$.

Let us examine the asymptotic behaviour of the thermodynamic
variables for large values of~$r_{\scriptscriptstyle\text{B}}$
or, equivalently, for small values
of~$u=1/r_{\scriptscriptstyle\text{B}}$.  The Maxwell
charge, ${\frak Q}=q$, and the entropy,
$S=2\pi\text{e}^{-2\phi_0}(2a^2-q^2)$, are both independent of the
size of the quasilocal system.  The quasilocal energy has the
following behaviour:
\begin{equation}
  E = \biggl\{ -\text{e}^{2\phi_0} u^{-1} + \biggl( a - \frac{q^2}{a}
  \biggr) + O(u) \biggr\} - E_{\scriptscriptstyle0} \;.
  \label{asymp electric energy}
\end{equation}
Unless $E_{\scriptscriptstyle0}$ is suitably chosen, the quasilocal
energy will diverge as~$r_{\scriptscriptstyle\text{B}}\to\infty$.
The temperature,
\begin{equation}
  T = \frac{1}{8\pi a}\, \biggl\{ \text{e}^{2\phi_0} + \biggl(
  a + \frac{q^2}{2a} \biggr)\,u + O(u^2) \biggr\},
  \label{asymp electric temperature}
\end{equation}
approaches the value~$\text{e}^{2\phi_0}/(8\pi a)$
as~$r_{\scriptscriptstyle\text{B}}\to\infty$, and the Maxwell
potential,
\begin{equation}
  {\frak V} = -\frac{q}{2a}\, \biggl\{ 1 + \text{e}^{-2\phi_0}
  ( q^2/2 - a^2 ) u + O(u^2) \biggr\},
  \label{asymp Maxwell potential}
\end{equation}
approaches~$-q/(2a)$.  The surface tension and the dilaton potential
have the following behaviour:
\begin{equation}
  {\cal S} = \frac{1}{8\pi}\, \biggl\{ \text{e}^{2\phi_0}u
  + \frac{q^2}{a}\,u^2 + O(u^3) \biggr\} + \biggl(
  \frac{\partial E_{\scriptscriptstyle0}}{\partial A} \biggr)
  \label{asymp surface tension}
\end{equation}
and
\begin{equation}
  \mu = \biggl\{ -2\text{e}^{2\phi_0}u^{-1} + \biggl(
  a - \frac{5q^2}{2a} \biggr) + O(u) \biggr\} + \biggl(
  \frac{\partial E_{\scriptscriptstyle0}}{\partial\phi} \biggr)
  \label{asymp dilaton potential}
\end{equation}
respectively.  As with the energy, the asymptotic behaviour of these
functions depends upon the choice of background action functional;
in particular, the dilaton potential will diverge
as~$r_{\scriptscriptstyle\text{B}}\to\infty$ unless a suitable choice
for~$E_{\scriptscriptstyle0}$ is made.  We will consider two possible
choices of~$I_{\scriptscriptstyle0}$ and calculate the asymptotic
behaviour of the energy, the surface tension, and the dilaton
potential.  (In the trivial case~$I_{\scriptscriptstyle0}=0$, we
see that the energy and the dilaton potential both diverge
as~$r_{\scriptscriptstyle\text{B}}\to\infty$.)

Our choice of~$I_{\scriptscriptstyle0}$ is
the functional of eq.~(\ref{background action}).  We find
$E_{\scriptscriptstyle0}=r_{\scriptscriptstyle\text{B}}
\text{e}^{-2\phi}$,
$\mu_{\scriptscriptstyle0}=2r_{\scriptscriptstyle\text{B}}
\text{e}^{-2\phi}$,
and~${\cal S}_{\scriptscriptstyle0}=-\text{e}^{-2\phi}/
(8\pi r_{\scriptscriptstyle\text{B}})$.  The quasilocal energy
is finite in the limit~$r_{\scriptscriptstyle\text{B}}\to\infty$
and approaches the value~$a$; the dilaton potential is also finite
and approaches~$q^2/(2a)$ in this limit; the surface tension,
${\cal S}\sim O(u^3)$ for large values
of~$r_{\scriptscriptstyle\text{B}}$.

\subsection{Thermodynamics of the Magnetic String Solution}

Recall now the magnetic string solution of Eqs.~(\ref{magnetic ghs}).
We wish to construct the thermodynamic variables in the same way
as we have just done for the electric string solution.  The
entropy and the energy can both be calculated as before.  They are
\begin{equation}
  S = 2\pi ar_{\scriptscriptstyle\text{B}}\text{e}^{-2\phi}\, \biggl(
  \frac{2a^2-q^2}{ar_{\scriptscriptstyle\text{B}}-q^2} \biggr)
  \label{magnetic entropy}
\end{equation}
and
\begin{equation}
  E = -\text{e}^{-2\phi}\,\biggl(
  f(r_{\scriptscriptstyle\text{B}})r_{\scriptscriptstyle\text{B}}
  +N(r_{\scriptscriptstyle\text{B}})\frac{q^2}{2a} \biggr)
  - E_{\scriptscriptstyle0}
  \label{magnetic energy}
\end{equation}
respectively.  We can also calculate the following intensive
variables as before:  The temperature is
\begin{equation}
  T = \frac{1}{8\pi aN(r_{\scriptscriptstyle\text{B}})},
  \label{magnetic temperature}
\end{equation}
the surface tension is
\begin{equation}
  {\cal S} = \frac{1}{8\pi r_{\scriptscriptstyle\text{B}}}
  \text{e}^{-2\phi}\,\biggl(
  f(r_{\scriptscriptstyle\text{B}})r_{\scriptscriptstyle\text{B}}
  +N(r_{\scriptscriptstyle\text{B}})\frac{q^2}{2a}
  +\frac{a}{N(r_{\scriptscriptstyle\text{B}})} \biggr)
  - {\cal S}_{\scriptscriptstyle0},
  \label{magnetic surface tension}
\end{equation}
and the dilation potential is
\begin{equation}
  \mu = -\text{e}^{-2\phi}\,\biggl(
  2f(r_{\scriptscriptstyle\text{B}})r_{\scriptscriptstyle\text{B}}
  +N(r_{\scriptscriptstyle\text{B}})\frac{q^2}{2a}
  +\frac{a}{N(r_{\scriptscriptstyle\text{B}})} \biggr)
  - \mu_{\scriptscriptstyle0}\;.
  \label{magnetic dilaton potential}
\end{equation}

There is a difficulty in evaluating the electromagnetic work term.
It is not possible to produce the Maxwell field strength of
eq.~(\ref{magnetic F}) with a non-singular potential~$\bbox{\frak A}$.
For example, the
potential~$\bbox{\frak A}=q(1-\cos\vartheta)\,\bbox{d}\varphi$ has
a Dirac string singularity at~$\vartheta=\pi$.  Because the terms in
eq.~(\ref{integral first law}) that give rise to the Maxwell work
terms involve projections of the potential, $\bbox{\frak A}$, it is
necessary to invoke some trick to avoid the Dirac string singularity
in the surface integral.  Let~$\delta Q_{\scriptscriptstyle\text{M}}$
denote the element of work from the Maxwell field.  From
eq.~(\ref{integral first law}), we have
\begin{equation}
  \beta\,\delta Q_{\scriptscriptstyle\text{M}} = \int_{\bar{\cal T}}
  N\,\bbox{d}\tau \wedge \bbox{\frak K}^a\,\delta{\frak W}_a
  \label{Maxwell work}
\end{equation}
where the ``${\frak V}\,\delta\bbox{\frak Q}$'' term does not appear
because $\bbox{\frak Q}$ vanishes for purely spatial Maxwell field
strength two-forms.  The topology of the Euclidean manifold,
$\bar{\cal M}$, is~$D^2\times S^2$ where $D^2$ is a two dimensional
disk and~$S^2$ is a two-sphere.  Consider instead the Euclidean
manifold~$\hat{\cal M}$ with topology~$D^2\times S_\theta^2$ where
$S_\theta^2$ is the fragment of a sphere $\vartheta\in[0,\theta)$
where~$\theta<\pi$.  The boundary of this manifold, $\hat{\cal T}$,
consists of the (periodic) history of the sphere fragment~$S_\theta^2$
of radius~$r_{\scriptscriptstyle\text{B}}$, and the (periodic) history
of the cone, $C_\theta^2$, of constant~$\vartheta=\theta$ and radius
running from the event horizon~$r_{\scriptscriptstyle\text{H}}$ to the
boundary~$r_{\scriptscriptstyle\text{B}}$.  Because the solution is
static, we find
\begin{equation}
  \beta\,\delta{\hat Q}_{\scriptscriptstyle\text{M}} = \Delta\tau
  \biggl( \int_{S_\theta^2} N\bbox{\frak K}^a\,\delta{\frak W}_a
  + \int_{C_\theta^2} N\bbox{\frak K}^a\,\delta{\frak W}_a \biggr) \;.
  \label{hat Maxwell work}
\end{equation}
Notice that the integral now avoids the Dirac string singularity.

Using eq.~(\ref{Maxwell current}) we see that~$\bbox{\frak K}^a=0$
on~$S_\theta^2$.  However, on~$C_\theta^2$, we have
\begin{equation}
  \bbox{\frak K}^\varphi = {}^2\!\bbox{\epsilon}\,
  \frac{\text{e}^{-2\phi}q}{4\pi r^3\sin\theta}
  \label{magnetic Maxwell current}
\end{equation}
and
\begin{equation}
  {\frak W}_\varphi = q(1-\cos\theta) \;.
  \label{magnetic Maxwell current potential}
\end{equation}
The volume element on~$C_\theta^2$ can be
written~${}^2\!\bbox{\epsilon}=-f^{-1}(r)\bbox{d}r\wedge
r\sin\theta\bbox{d}\varphi$ where the negative sign arises because the
unit normal, $n^a=r^{-1}(\partial/\partial\vartheta)^a$, points
inwards.  eq.~(\ref{hat Maxwell work}) can be computed:
\begin{equation}
  \beta\,\delta{\hat Q}_{\scriptscriptstyle\text{M}} = \beta\,\biggl(
  \frac{1-\cos\theta}{2} \biggr)\,\Biggl(
  \frac{q\text{e}^{-2\phi}}{f(r_{\scriptscriptstyle\text{B}})}\,
  \biggl( \frac{1}{r_{\scriptscriptstyle\text{B}}} -
  \frac{1}{r_{\scriptscriptstyle\text{H}}} \biggr)\Biggr) \delta q\;.
  \label{hat magnetic Maxwell work}
\end{equation}
Now we assume that
$Q_{\scriptscriptstyle\text{M}}=\lim_{\theta\to\pi}
{\hat Q}_{\scriptscriptstyle\text{M}}$.  Furthermore, we
write~$Q_{\scriptscriptstyle\text{M}}={\frak V}\,\delta{\frak Q}$
and, thus, define
\begin{equation}
  {\frak Q}=q
  \label{magnetic Maxwell charge}
\end{equation}
and
\begin{equation}
  {\frak V} = \text{e}^{-2\phi}\,
  \frac{q}{f(r_{\scriptscriptstyle\text{B}})} \biggl(
  \frac{1}{r_{\scriptscriptstyle\text{B}}} -
  \frac{1}{r_{\scriptscriptstyle\text{H}}} \biggr) \;.
  \label{magnetic Maxwell potential}
\end{equation}

The quasilocal energy can be written in terms of the extensive
variables $S$, $A$, $\phi$, and~$\frak Q$.  Then the relations
of eq.~(\ref{intensive variables}) can be shown to hold.  Thus, the
definition of the thermodynamic variables are consistent with the
first law of thermodynamics~(\ref{first law}).

The entropy, $S=2\pi\text{e}^{-2\phi_0}(2a^2-q^2)$, and
Maxwell charge, ${\frak Q}=q$, are independent of the size,
$r_{\scriptscriptstyle\text{B}}$, of the quasilocal region.
For~$r_{\scriptscriptstyle\text{B}}\to\infty$, $T\to(8\pi a)^{-1}$,
and~${\frak V}\to-\text{e}^{-2\phi_0}q/(2a)$.
Finally, we demonstrate the asymptotic behaviour of the energy, the
surface tension, and the dilaton force for small values
of~$u=1/r_{\scriptscriptstyle\text{B}}$.  We find
\begin{equation}
  E = \text{e}^{-2\phi_0}\,\biggl\{ u^{-1} + \biggl( a + \frac{q^2}{a}
  \biggr) + O(u) \biggr\} - E_{\scriptscriptstyle0}
  \label{asymp magnetic energy}
\end{equation}
\begin{equation}
  {\cal S} = \frac{1}{8\pi}\,\text{e}^{-2\phi_0}\, \biggl\{ u -
  \frac{q^2}{a}\,u^2 + O(u^3) \biggr\}
  + \biggl(\frac{\partial E_{\scriptscriptstyle0}}{\partial A}\biggr)
  \label{asymp magnetic surface tension}
\end{equation}
and
\begin{equation}
  \mu = \text{e}^{-2\phi_0}\,\biggl\{ -2u^{-1} + \biggl( a +
  \frac{5q^2}{2a} \biggr) + O(u) \biggr\}
  +\biggl(\frac{\partial E_{\scriptscriptstyle0}}{\partial\phi}\biggr)
  \label{asymp magnetic dilaton potential}
\end{equation}
respectively.  For the background action functional of
eq.~(\ref{background action}), we
have~$E_{\scriptscriptstyle0}=-\text{e}^{-2\phi}u^{-1}$.  Then the
quasilocal energy approaches the value~$\text{e}^{-2\phi_0}a$
as~$r_{\scriptscriptstyle\text{B}}\to\infty$, and the dilaton
potential approaches~$\text{e}^{-2\phi_0}\bigl(a+q^2/(2a)\bigr)$.  For
large~$r_{\scriptscriptstyle\text{B}}$, ${\cal S}\sim O(u^3)$.

\section{Summary}
\label{sec:summary}

We have shown that conformal invariance of conserved quantities
such as the mass is not automatically guaranteed: it crucially
depends on the choice of the background action functional.  We have
presented a background action functional that yields a conformally
invariant quasilocal mass for static, spherically symmetric
spacetimes, but it is not clear if this conformal invariance will
continue to hold for more general solutions.  This quasilocal mass
evaluated as~$r\to\infty$ for asymtotically flat spacetimes will
recover the {\sc adm} mass up to a constant factor.  
There is no guarantee that this conformally invariant mass 
is finite, as illustrated by the non-asymptotically flat solution
of Ref.~\cite{MZ96}. We have also     
presented an alternate choice of reference action functional based
on the one suggested by Hawking and Horowitz~\cite{HH95}.  However,
the resulting mass is not conformally invariant.

We have illustrated the calculation of the above masses for the
Garfinkle, Horowitz, and Strominger~\cite{GHS91,H92} static,
spherically symmetric string black hole solutions.
We found that the quasilocal mass
with our proposed background action functional is conformally
invariant, but that the mass arising from the action with the
reference given by the prescription of Hawking and Horowitz is not.
In addition,
we calculated the internal energy and other thermodynamic variables
for these solutions.  For the magnetically charged black hole
solution, we adopted a trick to avoid the Dirac string singularity.
The first law of thermodynamics is found to hold for these
thermodynamic variables.  Although the individual thermodynamic
variables depend on the choice of the background action
funtional, the first law of thermodynamics does not.  Also, the first
law of thermodynamics is conformally invariant even though the
individual thermodynamic variables are not.

\acknowledgments

This work has been supported in part by the Natural Sciences and
Engineering Research Council (NSERC) of Canada.

\end{document}